\documentclass[preprint,12pt]{elsarticle}

\usepackage[colorlinks=false,citecolor=black,linkcolor=magenta]{hyperref}
\usepackage{geometry}
\usepackage{booktabs}
\usepackage{enumitem}
\usepackage{natbib}
\usepackage{algpseudocode}
\usepackage{algorithm}
\usepackage{array}
\usepackage{rotating}
\usepackage{graphicx}
\usepackage{multirow}
\usepackage{setspace}
\usepackage{subcaption}
\usepackage{amsmath}
\usepackage{threeparttable}
\usepackage{color}
\usepackage{amssymb}
\usepackage{breqn}
\usepackage{lscape}
\usepackage[normalem]{ulem}
\usepackage{soul}

\newcolumntype{P}[1]{>{\centering\arraybackslash}p{#1}}
\newcolumntype{Q}[1]{>{\raggedleft\arraybackslash}p{#1}}
\newcolumntype{C}[1]{>{\centering\arraybackslash}p{#1}}
\graphicspath{{./graphics/}}

\journal{Sustainable Cities and Society}
\usepackage{lineno}

\usepackage{xcolor}
\renewcommand\fbox{\fcolorbox{gray!25}{white}}
\makeatletter
\def\ps@pprintTitle{%
 \let\@oddhead\@empty
 \let\@evenhead\@empty
 \def\@oddfoot{}%
 \let\@evenfoot\@oddfoot}

\makeatother

\begin{document}

\begin{frontmatter}
\title{InfraRisk: An Open-Source Simulation Platform for Asset-level Resilience Analysis in Interconnected Infrastructure Networks}


\author{Srijith Balakrishnan, Ph.D.\corref{correspondingauthor}}
\address{Postdoctoral Researcher, Future Resilient Systems, Singapore-ETH Centre at CREATE, 1~Create Way, Singapore 138602}
\author{Beatrice Cassottana, Ph.D.\corref{cor3}}
\address{Postdoctoral Researcher,  Future Resilient Systems, Singapore-ETH Centre at CREATE, 1~Create Way, Singapore 138602}
\cortext[correspondingauthor]{Corresponding author. Email: \href{mailto:srijith.balakrishnan@sec.ethz.ch}{srijith.balakrishnan@sec.ethz.ch}, Tel: +65 98063679}

\thispagestyle{empty}
\begin{abstract}

Integrated simulation models are emerging as an alternative for analyzing large-scale interdependent infrastructure networks due to their modeling advantages over traditional interdependency models. This paper presents an open-source integrated simulation package for the asset-level analysis of interdependent infrastructure systems. The simulation platform, named `InfraRisk' and developed in Python, can simulate disaster-induced infrastructure failures and subsequent post-disaster restoration in interconnected water-, power-, and road networks. InfraRisk consists of an infrastructure module, a hazard module, a recovery module, a simulation module, and a resilience quantification module. The infrastructure module integrates existing infrastructure network packages (\textit{wntr} for water networks, \textit{pandapower} for power systems, and a static traffic assignment model for transportation networks) through an interface that facilitates the network-level simulation of interdependent failures. The hazard module generates infrastructure component failure sequences based on various disaster characteristics. The recovery module determines repair sequences and assigns repair crews based on predefined heuristics-based recovery strategies or model predictive control (MPC) based optimization. Based on the schedule, the simulation module implements the network-wide simulation of the consequences of the disaster impacts and the recovery actions. The resilience quantification module offers system-level and consumer-level metrics to quantify both the risks and resilience of the integrated infrastructure networks against disaster events. InfraRisk provides a virtual platform for decision-makers to experiment and develop region-specific pre-disaster and post-disaster policies to enhance the overall resilience of interdependent urban infrastructure networks.

\end{abstract}


\end{frontmatter}
\par





\newpage

\section{Introduction}
Climate-induced extreme weather events and emerging threats, such as cyber-attacks and acts of terrorism, pose severe risks to the uninterrupted functioning of urban infrastructure networks. Of significant concern are the short- and long-term societal and economic losses resulting from infrastructure service disruptions, which may be compounded by the interdependencies among critical infrastructure networks. This has been evident in the recent 2021 Texas snowstorm, which caused power blackouts and water supply disruptions in the industrial hubs of Central Texas \cite{Busby2021}, leading to ripple effects in the national and global supply chains. Similar events in the past have shown that the scale of the indirect economic impacts resulting from infrastructure disruptions are of magnitude larger than the direct costs incurred due to infrastructure failure and restoration \cite{NRC1999,Hallegatte2019}.

Empirical data and simulation studies confirm that judicious interventions to improve resilience against infrastructure failures resulting from external factors could not only reduce operational disruptions to individual infrastructure systems but also decrease the likelihood of large-scale network-wide consequences \cite{DeAlmeida2016,Sadashiva2021}. Such interventions may be classified according to the targeted resilience characteristic(s) of the infrastructure system, namely resistant capacity, absorptive capacity, and restorative capacity \cite{Vugrin2011,Ouyang2014}.

However, the challenge is to identify and evaluate various feasible resilience alternatives (both system-level and asset-level) and implement the cost-effective ones \cite{Najarian2020}. Moreover, such an effort requires frameworks and tools for a holistic understanding of the network-wide consequences of various system-level and asset-level resilience strategies \cite{Ouyang2014,Setola2016} so that the cost and benefit aspects can be evaluated. In addition, the increasing societal and economic significance in infrastructure resilience studies also strengthens the case for developing simulation tools with the capability to perform more granular analyses.

While modeling and simulation tools have been extensively applied for resilience analysis in individual infrastructure systems, those intended for integrated infrastructure systems for asset-level analyses are very few and require further research \cite{Saidi2018}. In this study, an open-source simulation package for the analysis of interdependent infrastructure systems is presented. The simulation platform is named `InfraRisk' and is developed in Python. The platform is capable of simulating and evaluating disaster-induced infrastructure failures and subsequent post-disaster restoration in interconnected water-, power-, and road networks.

The rest of the paper is organized as follows: Section 2 provides a brief overview of the existing tools for interdependent infrastructure simulation; Section 3 discusses the architecture of InfraRisk simulation platform and its various functionalities; Section 4 presents an experimental simulation of interdependent effects of a few disaster scenarios on a test infrastructure network using the InfraRisk platform, and Section 5 summarizes the conclusions.

\section{Related Work}
While disaster resilience has been historically acknowledged as an indispensable quality of infrastructure systems, the need for considering their (inter)dependencies in assessing their ability to withstand unanticipated events has gained attention only in the last three decades. Until then, the focus was on developing best practices to enhance the physical security and robustness of individual infrastructure systems \cite{Fisher2018}. However, several disasters in the past, such as the World Trade Center attack in 2001, the Northeast blackout of 2003, the Indian Ocean earthquake and tsunami in 2004, and Hurricane Katrina in 2005, showed that the interconnections among critical infrastructure sectors intensified the societal and economic loss incurred by disasters \cite{ORourke007}. Over the years, several countries incorporated infrastructure systems and their interdependencies as major considerations in aspects concerning national security and extended support for research and development in the area of infrastructure resilience \cite{EU2006,TheWhiteHouse2013}. Consequently, academic interest in the identification, classification, and modeling of infrastructure interdependencies has grown considerably in the recent past, leading to the development of numerous frameworks, methods, and models for the analysis of interdependent infrastructure networks.

\subsection{Traditional infrastructure simulation models}
Traditional infrastructure interdependency models can be categorized into two groups: empirical models and computational models \cite{Mitsova2021}. Empirical models rely on data related to historical infrastructure breakdowns for characterizing the interactions between infrastructure systems. Empirical models could be built using qualitative data (for example, based on expert judgments \cite{Cooke2004,Zorn2021}) or quantitative data (for example, based on post-disaster impact assessment reports, newspaper reports, and official incident reports \cite{McDaniels2007,Luiijf2009}). 
The major limitation of empirical models is their incapability to identify possible failure and cascading event scenarios that may not yet have occurred.

On the contrary, the computational approach explicitly captures physical, cyber, geographic, and logical relationships among infrastructure systems using mathematical and logical functions along with functional data and attempts to mimic real-world systems in a controlled environment. Therefore, such models are effective in understanding the role of infrastructure properties and resilience interventions in system response. The most common computational models are graph-based models, system-dynamics models, and agent-based models. 

In graph-based models, abstractions of infrastructure systems are created by representing infrastructure components producing or consuming resources using nodes and their interrelationships using edges \cite{Svendsen2007,Svendsen2008}. Graph theory and network-flow optimization concepts have been extensively applied to study the impacts of infrastructure component failures and their consequences \cite{Praks2017,Holden2013}. The graph-based models do not consider detailed functional aspects of network components. 

System dynamics-based models use the concepts of feedback loops, stocks, and flows to create abstractions of integrated infrastructure networks. System dynamics-based models simulate the dynamic and evolutionary effects in the infrastructure system by testing various policies and investment alternatives using stock and flow variables \cite{Powell2008,Pasqualini2005}. The notable drawbacks of system dynamics models are their incapability to analyze infrastructure systems at component-level, the need for a large amount of data for calibration, and excessive dependence on expert judgments for establishing feedback loops in the model \cite{Ouyang2014}.

The agent-based modeling (ABM) is a bottom-up approach for modeling integrated infrastructure networks at a more granular scale by considering them as system-of-systems (SoS) \cite{Nilsson2006}. ABMs simulate infrastructure components as agents and allow them to interact with each other in representative operating environments to obtain quantitative insights into the system behavior \cite{Tesfatsion2003,Thompson2019,Balakrishnan2020a}. Some of the key issues pertaining to ABM are its over-dependence on assumptions, calibration of parameters, and validation of the results \cite{Ouyang2014}.

In addition to empirical and computational models, economic theory models, such as input-output models \cite{Haimes2005,Oliva2010}, have also been extensively used for modeling interdependencies. However, such models, like empirical models, depend on disaster data for calibrating the strength of dependencies among infrastructure sectors.

\subsection{Integrated infrastructure simulation models}
As an alternative to traditional interdependency models, several researchers have proposed the idea of integrating existing infrastructure-specific simulation models \cite{Nan2011,Eusgeld2011,Portante2017}. Such integrated models intend to take full advantage of existing domain-specific models to simulate the global behaviour of complex heterogeneous systems \cite{Gomes2018}. The key challenges of developing an integrated infrastructure model are time synchronization and coupling of constituent simulators having divergent functional dynamics, complexity, and time-scale \cite{Caire2013}.

As far as critical infrastructure systems are concerned, integrated models have been used in various fields. For example, \cite{Monti2009} developed an integrated model by combining VTBPro (power system), Matlab Simulink (control system) and OpNet (communication system) simulation tools for studying the role of communication networks in the performance of power grids. In another study, \cite{Erdener2014} developed an electricity and gas network simulation and analysis platform by integrating infrastructure specific simulation models using a MATLAB interface.

A more advanced set of integrated models, known as co-simulation models, adopt a distributed simulation approach. Co-simulation frameworks can be used to operate computationally intensive domain-specific submodels on different computers and seamlessly interact with each other on a real-time basis, enhancing both efficiency and usability \cite{Wang2022}. Several frameworks and standards have been developed to implement co-simulation of complex systems, such as functional mock-up interface (FMI), high-level architecture (HLA), and distributed interactive simulation (DIS) \cite{Ahmad2016}. In recent years, the high-level architecture (HLA) standard has been increasingly adopted to develop interdependent infrastructure models for perform simulations in a distributed environment \cite{Dubaniowski2019,Wang2022}.

Recent studies have also extended the scope of integrated infrastructure models by combining infrastructure-specific simulators with other urban system models for investigating various aspects of disaster resilience of urban regions. For example, \cite{Marasco2021} developed a Python-based integrated platform for assessing the vulnerability and resilience of urban power and water networks against seismic events by combining infrastructure simulators and agent-based socio-technical models. Similarly, \cite{Yang2021} combined geographic information system (GIS) tools, building information modeling (BIM) tools, and domain-specific infrastructure simulation models to develop an integrated model to analyze the risks of urban flooding on both infrastructure and communities.

\subsection{Research gaps}
Most of the above traditional interdependent infrastructure simulation models provide a system-level abstraction of infrastructure network performance with limited emphasis on the asset-level operational characteristics. Moreover, due to their intrinsic limitations and homogeneous modeling approaches, traditional models are incapable of leveraging upon the existing domain knowledge in the constituent infrastructure sectors. Consequently, their ability to perform analyses involving realistic resilience interventions is often limited. Recent advancements in integrated modeling techniques have led to its greater adoption in interdependent infrastructure modeling and analysis. However, the use of integrated models in infrastructure analysis is an emerging field and in the early stages of adoption, especially in the case of civil infrastructure systems. For the same reason, there is a lack of open-source integrated infrastructure tools for analyzing interdependent infrastructure systems such as power, water, and transportation networks. An open-source integrated platform to simulate disaster impacts on large-scale interdependent infrastructure systems could be of interest to a wide range of stakeholders involved in urban resilience decision-making.

\section{Methodology}

\subsection{Methodological Framework}
\begin{sidewaysfigure}[htbp]
\centering
\includegraphics[width=20cm,clip]{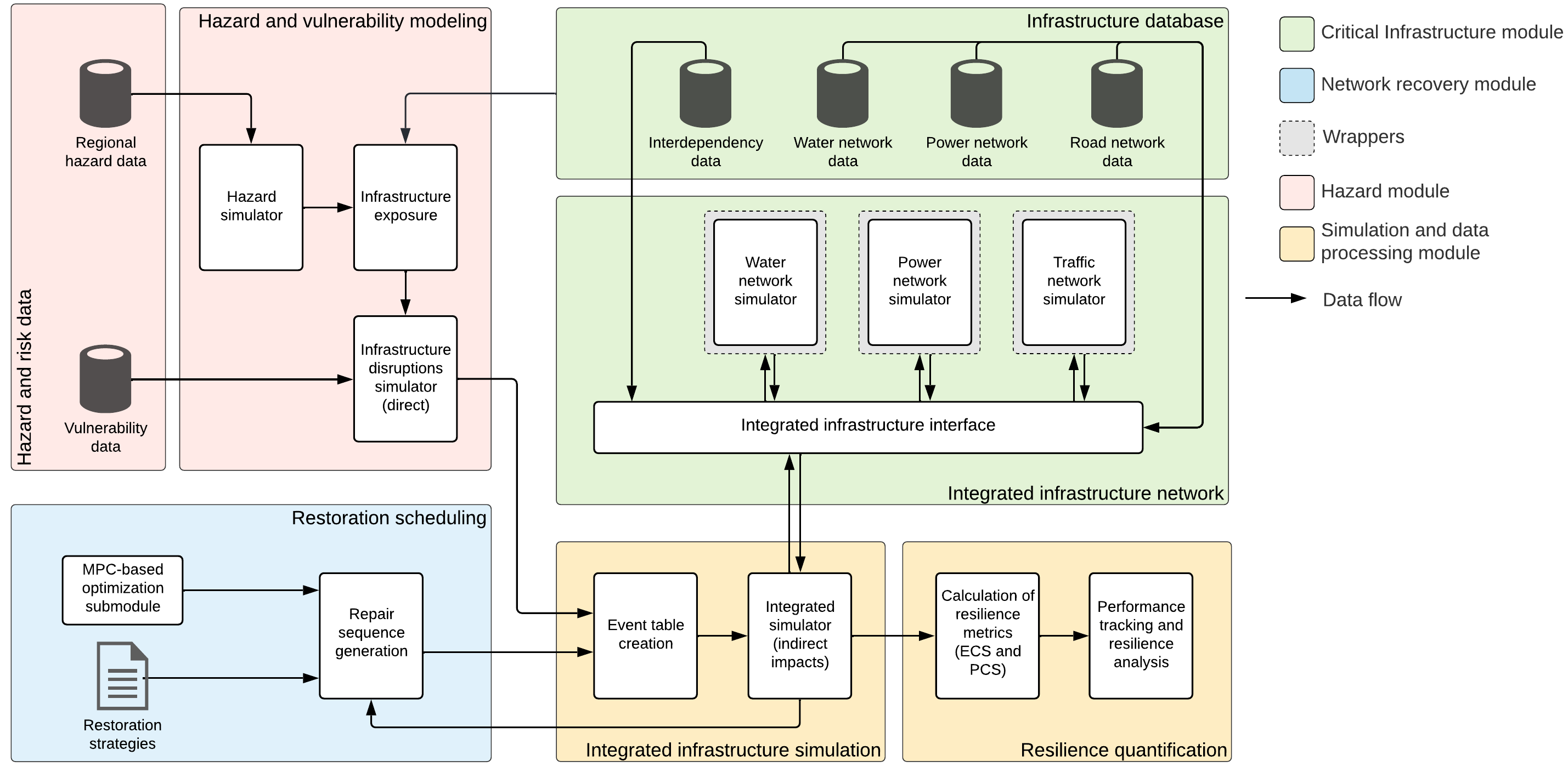}
\caption{Software implementation framework adopted in the InfraRisk simulation platform}
\label{fig:framework}
\end{sidewaysfigure}

Figure~\ref{fig:framework} illustrates the methodological framework adopted in the InfraRisk simulation platform. The package is based on the widely accepted risk- and resilience analysis framework \cite{Argyroudis2020,Balakrishnan2020}. In this framework, the most important component is an interdependent infrastructure network model that consists of various infrastructure systems of interest. In addition, the major hazards in the region can also be modeled. Further, the vulnerabilities in the networks to those hazards are mapped and the direct impacts (physical and functional failures in the infrastructure networks) are simulated using the hazard model. For scheduling post-disaster restoration/repair actions, a recovery model is also developed.  The restoration actions are prioritized based on specific recovery strategies or optimization methods. Therefore, the indirect failures in the network are simulated using the interdependent infrastructure model based on the initial failure events and the subsequent repair actions. The asset- and system-level operational performance are tracked using appropriate resilience metrics. 

\subsection{Implementation in Python}
Python programming language is chosen for developing InfraRisk because of its versatility, ease of use, and extensive open-source libraries. Python repositories house several efficient libraries to model individual infrastructure systems, such as water (\textit{wntr}) and power (\textit{pypower} and \textit{pandapower}) systems, which could be adopted as domain-specific infrastructure simulators. 

The basic idea behind the InfraRisk simulation package is to integrate existing infrastructure-specific simulation models through an object-oriented interface so that interdependent infrastructure simulation can be achieved. Interfacing requires identifying and modeling the dependencies among various infrastructure assets and time-synchronization among infrastructure simulation models. To address the above challenges, InfraRisk is built using a sequential simulation framework (Figure~\ref{fig:framework}). The advantage of this approach is that it simplifies the efforts for data preparation and enables the complete utilization of asset-level modeling features of the domain-specific infrastructure models.

InfraRisk consists of five modules, namely, (a) integrated infrastructure network simulation, (b) hazard initiation and vulnerability modeling, (c) recovery modeling, (d) simulation of direct and indirect effects, and (e) resilience quantification. In the rest of the section, a detailed discussion on each of the above modules is provided.

\subsubsection{Integrated infrastructure network simulation}
This module houses the three infrastructure network models to simulate power-, water-, and transportation networks. These models are developed using existing Python-based packages. In order to model the power network, \textit{pandapower} is employed \cite{Thurner2018}. The water distribution network is modeled using \textit{wntr} package \cite{Klise2020}. The traffic network model provides the travel costs for traveling from one point in the network to another and is modeled using the static traffic assignment method \cite{Boyles2020}. All three packages have network-flow optimization models that identify the steady-state resource flows in the network considering the operational constraints. The details of the packages are presented in Table~\ref{tab:infra_packages}.

\begin{table}[htbp]\small
	\caption{Infrastructure packages using in the simulation model}
	\begin{tabular}{p{2.5cm}p{2.5cm}p{8cm}}
	\toprule
	Infrastructure & Package & Capabilities\\ 
	\midrule
	Power & \textit{pandapower} & 
	\begin{minipage}[t]{\linewidth}\begin{itemize}[leftmargin=6pt,noitemsep,nolistsep,after=\strut]
        \item Capable of generating power networks with standard components such as lines, buses, and transformers based on design data.
		\item Capable of performing power-flow analysis.
    \end{itemize}\end{minipage} \\
	\midrule
	Water & \textit{wntr} & \begin{minipage}[t]{0.6\textwidth}
		\begin{itemize}[leftmargin=6pt,noitemsep,nolistsep,after=\strut]
			\item Capable of generating water networks with standard components such as pipes, tanks, and nodes based on design data.
			\item Capable of performing pressure dependent demand or demand-driven hydraulic simulations.
		\end{itemize}
	\end{minipage} \\
	\midrule
	Transportation & static traffic assignment package & 
	\begin{minipage}[t]{0.6\textwidth}
		\begin{itemize}[leftmargin=6pt,noitemsep,nolistsep,after=\strut]
			\item Capable of implementing static traffic assignment and computing travel times between origin-destination pairs.
		\end{itemize}
	\end{minipage} \\
	\bottomrule
	\end{tabular}
	\label{tab:infra_packages}
\end{table}

The \textit{pandapower} package can be used to determine the steady-state optimal power flow for a given set of network conditions. The optimal power flow problem, solved by \textit{pandapower}, attempts to minimize the total power distribution costs in the network under load flow-, branch-, bus-, and operational power constraints (Equation~\ref{eq:pandapowereqs}) \cite{Thurner2018}.

\begin{equation}
    \begin{aligned}
        \text{min} \quad & \sum_{i\in {gen,sgen,load,extgrid}}P_{i}\times f_{i}\left (P_{i}  \right )\\
        \textrm{s.t.} \quad & P_{min, i}\leq P_{i}\leq P_{max,i} & \forall i \in {gen,sgen,extgrid,load}\\
            & Q_{min, i}\leq Q_{i}\leq Q_{max,i} & \forall i \in {gen,sgen,extgrid,load}\\
            & V_{min,j} \leq V_{j}\leq V_{max,j} & \forall j\in {bus}\\
            & L_{k} \leq L_{max,k} & \forall k \in {trafo,line,trafo3w}
    \end{aligned}
    \label{eq:pandapowereqs}
\end{equation}
where $i$, $j$, and $k$ are the power system components, $gen$ is the set of generators, $sgen$ is the set of static generators, $load$ is the set of loads, $extgrid$ is the set of external grid connections to the network, $bus$ is the set of bus bars, $trafo$ is the set of transformers, $line$ is the set of lines, and $trafo3w$ is the set of three winding transformers, $f_{i}(\cdot)$ is the cost function, $P_{i}$ is the active power in $i$, $Q_{i}$ is the reactive power in $i$, $V_{j}$ is the voltage in $j$ and $L_{k}$ is the loading percentage in $k$.

The \textit{wntr} package can simulate water flows in water distribution networks using two common approaches, namely, demand-driven analysis (DDA) and pressure-dependent demand analysis (PDA). While DDA assigns pipe flows based on the demands irrespective of the pressure at demand nodes, PDA assumes that the demand is a function of the pressure at which water is supplied. The PDA approach is more suitable for pressure-deficient situations, such as disaster-induced disruptions to water infrastructure. In the case of PDA, the actual node demands is computed as a function of available the water pressure at the nodes as in Equation~\ref{eq:PDA} \cite{Klise2020}.

\begin{equation}
    d_{i}(t) = \begin{cases}
    0 & p_{i}(t) \leq P_{0} \\
    D_{i}(t)\left (\frac{p_{i}(t) - P_{0}}{P_{f} - P_{0}}  \right )^{\frac{1}{e}} & P_{0} < p_{i}(t) \leq P_{f}\\
    D_{i} & p_{i}(t) > P_{0}
    \end{cases}
    \label{eq:PDA}
\end{equation}
where $d_{i}(t)$ is the actual demand at node $i$ at time $t$, $D_{i}(t)$ is the desired demand at a node $i$ at $t$, $p_{i}(t)$ is the available pressure in node $i$ at $t$, $P_f$ is the nominal pressure, and $P_0$ is the lower pressure threshold,  below which no water is consumed.

The traffic conditions on the road network is modeled using the static traffic assignment method based on the principle of user-equilibrium. Under user-equilibrium, every user tries to minimize their travel costs. The traffic assignment problem considered in InfraRisk package is formulated as follows (Equation~\ref{eq:staeqs}) \cite{Boyles2020}.

\begin{equation}
    \begin{aligned}
        \min_{\mathbf{x,h}} \quad & \sum_{(i,j)\in A} \int_{0}^{x_{ij}} t_{ij}(x_{ij})dx\\
        \textrm{s.t.} \quad & x_{ij} = \sum_{\pi \in \Pi} h^{\pi}\delta_{ij}^{\pi} & \forall (i,j) \in A\\
        & \sum_{\pi \in \Pi^{rs}} h^{\pi} = d^{rs} & \forall (r,s) \in Z^{2}\\
        & h^{\pi} \geq 0 & \forall \pi \in \Pi
    \end{aligned}
    \label{eq:staeqs}
\end{equation}
where $A$ is the set of all road links with $i$ and $j$ as the tail and head nodes, $t_{ij}$ is the travel cost on link $(i,j)$, $x_{ij}$ is the traffic flow on link $(i,j)$, $h^{\pi}$ is the flow on path $\pi \in \Pi$, $\delta_{ij}^{\pi}$ is an indicator variable that denotes whether $(i,j)$ is part of $\pi$, $d^{rs}$ is the total flow between origin-destination pair $r,s$.

The module also consists of an interdependency layer which serves as an interface between infrastructure network pairs. The interdependency layer stipulates the different pieces of information that can be exchanged among individual infrastructure networks and their respective formats. The interdependency submodule also stores information related to the various component-to-component couplings between infrastructure pairs. The module facilitates the communication between infrastructure systems and enables information transfer triggered by dependencies. Currently the following dependencies are considered.

\begin{itemize}
\item Power-water dependencies, which include dependency of water pumps on electric motors and generators on reservoirs (hydro-power).
\item Dependencies also exist between traffic network and the other two infrastructure models, as the former provides access to the latter. The disruptions to transportation infrastructure components and their recovery are key considerations that influence the restoration and recovery of all other infrastructure networks. The module also stores the functional details of all network components, including their operational status after a disaster.
\end{itemize}

The interdependency layer communicates with the infrastructure simulators through inbuilt functions (wrappers).

\subsubsection{Hazard initiation and vulnerability modeling}
\label{sec:hazards}
The hazard module generates disaster scenarios and initiates disaster-induced infrastructure failures based on their vulnerability. The hazard initiation and the resulting infrastructure asset failures is the first step in the interdependent infrastructure simulation. The probabilistic failure of an infrastructure component is modeled as follows (Equation~\ref{eq:failureprob}):

\begin{equation}
    p\left ( \text{failure}_{i} \right ) = p\left (\text{hazard}  \right ) \times p\left ( \text{exposure}_{i}|\text{hazard} \right ) \times p\left ( \text{failure}_{i}| \text{exposure}_{i} \right )
    \label{eq:failureprob}
\end{equation}
where $i$ is the component, $p(\cdot)$ is the probability. The probability of failure of a component is computed as the product of the probability of the hazard, the probability of the component being exposed to the hazard if it occurs, and the probability of failure of the component if it is exposed to the event.

The hazard module generates infrastructure component failures based on the hazard characteristics. In its current version, point events (such as random failures, terrorist attacks, and fire incidents), track-based events (such as floods, cyclones, and tornadoes), and random events can be generated using the module. 

For generating point events, three parameters, namely, the location of occurrence of the event, the radius of the impact, and the intensity of the event are to be provided. The hazard module allows five predefined intensity levels-- low, moderate, high, extreme, and random. The conditional failure probability ($P( \text{failure}_{i}| \text{exposure}_{i})$) is dependent on the intensity level. The conditional probability of a component being exposed to the event ($P( \text{exposure}_{i}|\text{hazard}$) is defined as a function of the ratio of the distance of the component to the location of occurrence of the event and the radius of impact.

In the case of track-based events, the input parameters are the track of the event, intensity of the event, and the offset distance of impact. In the case of flood events, the GIS data of the regional water bodies are also needed. For other track-based events, such as tornadoes and hurricanes, the tracks are randomly generated using a spline function. For all track-based events, the conditional probability of being exposed to the event is calculated as a function of the ratio of the perpendicular distance from the component to the track to the offset distance of impact.

The third type of disruptions that can be generated is the random failure events. The user can choose the number of components that needs to be randomly failed in each disaster scenario.

\begin{figure}[h!]
  \begin{subfigure}[t]{.49\textwidth}
    \centering
    \fbox{\includegraphics[width=\linewidth, trim = {1.5cm 1cm 3cm 1cm}, clip]{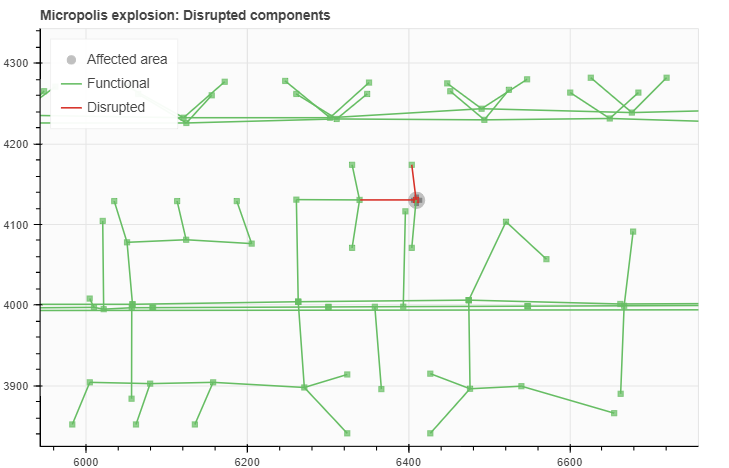}}
    \caption{Random event: power line failure}
  \end{subfigure}
  \hfill
  \begin{subfigure}[t]{.49\textwidth}
    \centering
    \fbox{\includegraphics[width=\linewidth, trim = {1.5cm 1cm 3cm 1cm}, clip]{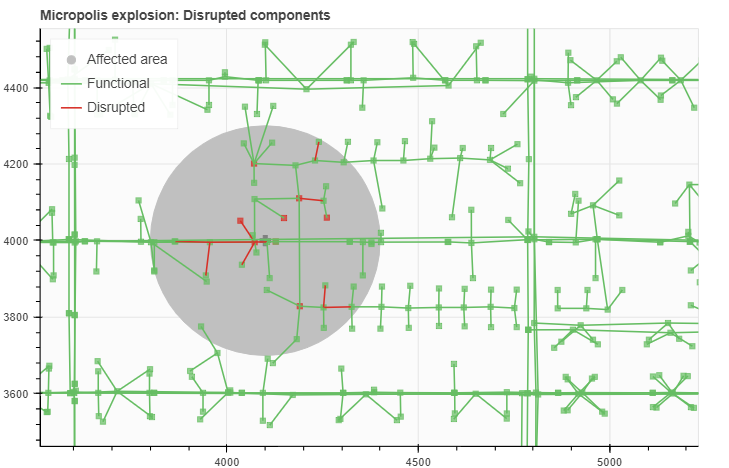}}
    \caption{Point event: explosion}
  \end{subfigure}

  \medskip

  \begin{subfigure}[t]{.49\textwidth}
    \centering
    \fbox{\includegraphics[width=\linewidth, trim = {1.5cm 1cm 3cm 1cm}, clip]{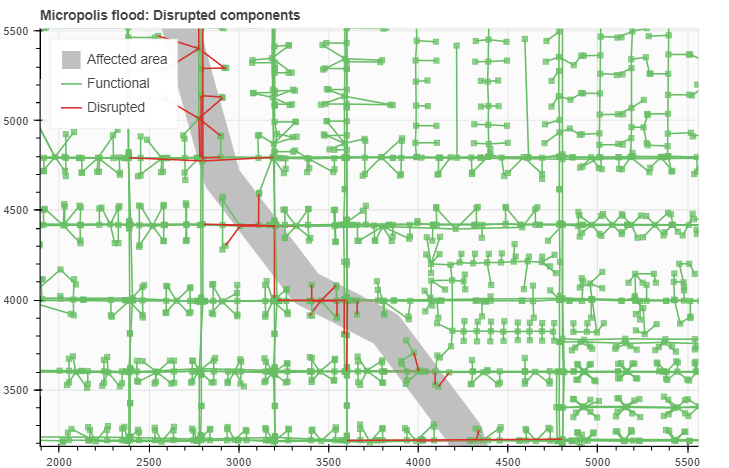}}
    \caption{Track-based event: flood}
  \end{subfigure}
  \hfill
  \begin{subfigure}[t]{.49\textwidth}
    \centering
    \fbox{\includegraphics[width=\linewidth, trim = {1.5cm 1cm 3cm 1cm}, clip]{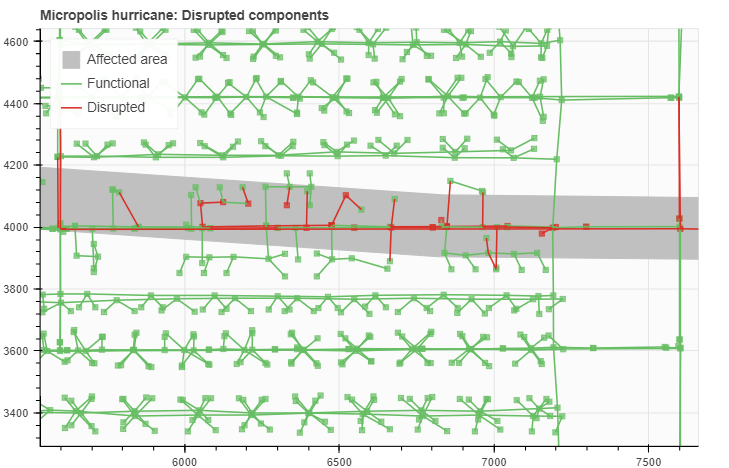}}
    \caption{Track-based event: tornado}
  \end{subfigure}
  \caption{Examples of disaster scenarios and resultant infrastructure failures generated by the hazard module}
  \label{fig:disasters}
\end{figure}

A few disaster scenarios generated by the hazard module are illustrated in Figure~\ref{fig:disasters}. The radius/offset distance of impact is marked in grey color, whereas the failed components as determined by Equation~\ref{eq:failureprob} is denoted using red color.

After a disaster scenario is generated, the module stores the details of infrastructure disruptions (disrupted components, time of disruption, and the severity of disruption) in the local directory for use during the network simulation.

\subsubsection{Recovery modeling}
The third module, which is the recovery module, determines how the repair actions are sequenced and implemented. The two major factors that influence recovery are the availability of repair crews and the criteria used for selecting subsequent components to restore. In InfraRisk, the user can specify the number of crews deployed for restoration of the three infrastructure networks and their initial locations. 

In the current simulation model, the repair sequence can be derived using two approaches. The first approach is to adopt repair strategies based on performance- and network-based heuristics for developing repair sequences. Currently, there are four inbuilt strategies based on the following criteria: 
\begin{enumerate}
    \item Maximum flow handled: The resource-flow during normal operating conditions could reflect the importance of an infrastructure component to the network. The maximum resource-flow handled by a component, considering the temporal fluctuations, can be used as a performance-based heuristics to prioritize failed components for restoration.
    \item Betweenness centrality: Centrality is a graph-based measure that is used to denote the relative importance of components (nodes and links) in a network. Betweenness centrality is often cited as an effective measure to identify critical infrastructure components \cite{Almoghathawi2019}.
    \item Crew distance: In large networks, logistics may be an issue for performing repair actions. Repairing the nearest failed components and advancing to the farthest ones save time and effort for logistics.
    \item Landuse/zone: Certain regions in a network may have consumers with large demands or critical to the functioning of the whole city. Industrial zones and central business districts are critical from both societal and economic perspectives.
\end{enumerate}
While it is comparatively easier to derive repair sequences based on heuristics, they may not guarantee optimal recovery of the network.

The second approach is an optimization model leveraging on the concept of model predictive control (MPC) \cite{Camacho2007}. In this approach, first, out of $n$ repair steps, the solution considering only $k$ steps (called the prediction horizon) is computed. Next, the first step of the obtained solution is applied to the system and then the process is repeated for the remaining $n-1$ components until all components are scheduled for repair. In the context of the integrated infrastructure simulation, the optimizer module evaluates repair sequences of the length of the prediction horizon for each infrastructure (assuming that each of the infrastructure has a separate recovery crew) based on a chosen resilience metric \cite{Kottmann2021}. The optimal repair sequence is found by maximizing the resilience metric. At this stage, the optimal repair action in each prediction horizon is computed using a brute-force approach where the resilience metric is evaluated for each of the possible repair sequences. The major limitation of MPC is that it is suitable only for small disruptions involving a few component failures; MPC becomes computationally expensive to derive optimal restoration sequences for larger disruptions due to the large number of repair permutations it has to simulate.

\subsubsection{Simulation of direct and indirect effects}
The simulation module implements the integrated infrastructure simulation in two steps, namely, event table generation and interdependent infrastructure simulation.The objective of the event table is to provide a reference object to schedule all the disruptions and repair actions for implementing the interdependent network simulation.

\begin{figure}[htbp]
\centering
    \includegraphics[width=14cm,clip]{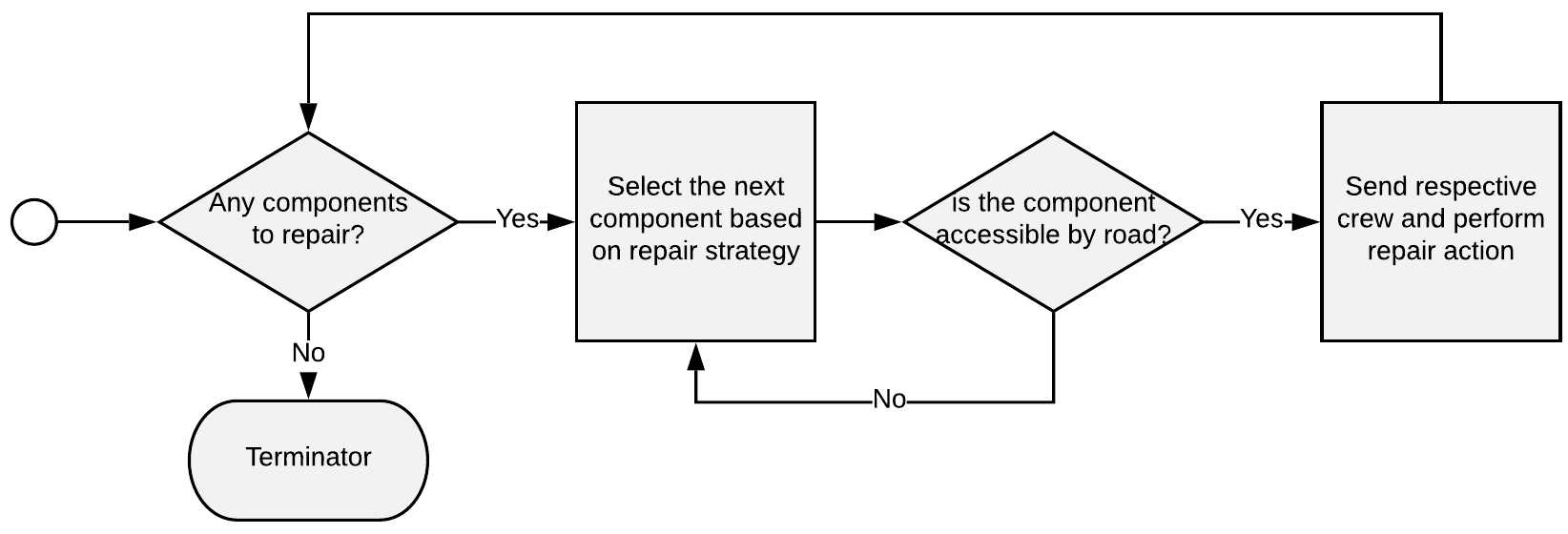}
    \caption{Algorithm used for scheduling the repair actions scheduled using a predefined recovery strategy}
    \label{fig:algorithm}
\end{figure}

While individual infrastructure agencies may have finalized a restoration plan immediately after a disaster (especially one based on the strategies defined by heuristics), there are several factors that may influence its implementation. The most important hurdle is accessing the locations of disrupted components, as roads may also be disrupted either due to damage or debris \cite{Comes2014,Iloglu2020}. In such a scenario, the crews may modify the repair sequence by prioritizing those components which are accessible under prevalent road network conditions. The modified repair sequence considering road access and the corresponding repair  start and end times are computed by the model using the algorithm illustrated in Figure~\ref{fig:algorithm}.

The algorithm first checks if there are any components affected by the disruptive event. If there are any, the infrastructure repair sequences based on the strategy chosen by the user is derived and the repair tasks are assigned to the respective infrastructure network crews. Now, for each of the infrastructure system, one component at a time is selected in the order and its accessibility by the repair crew from their current location is ascertained. If the component is accessible, the shortest path (based on the static traffic assignment method under prevailing road conditions) is identified and the travel time is computed. Next, the start and the end times of the repair action are computed based on the travel time and the component repair time. At the end of the repair, if the component belongs to the transportation network, the static traffic assignment model is rerun and the network flows and travel times are updated. This completes one cycle of the algorithm and the same process is repeated until all the components are scheduled for repair.

The component failures, repair actions, and the respective time-stamps, as obtained from the algorithm described in Figure~\ref{fig:algorithm}, are recorded in an event table for later use in the simulation module. The simulation platform uses the event table as a reference to modify the operational status of network components during the simulation, so that the consequences of disaster events and repair actions are reflected in the network performance. The recovery module also stores details including the number of repair crews for every infrastructure network, and their initial locations.

Once the event table is created, the next step is to simulate the interdependent effects resulting from the component disruptions and the subsequent restoration efforts. One of the main challenge in simulating the interdependent effects using a platform that integrates multiple infrastructure models is the time synchronization. Time synchronization is essential in InfraRisk for the following reasons:
\begin{itemize}
    \item The water network model (\textit{wntr}) simulates the water flows in the network for specified intervals, whereas the power system model (\textit{pandapower} computes power flows for a given time. 
    \item The network-wide effects are immediately reflected in power systems, whereas, in the case of water networks, there might be delayed responses, such as gradual emptying of the tanks, which need to be captured by the resilience metrics.
\end{itemize}

In order to synchronize the times, the power- and water- network models are run successively for every subsequent time-intervals in the event table. The required water network metrics are collected for every one minute of simulation time from the \textit{wntr} model, whereas power network characteristics at the start of every time interval is recorded from the \textit{pandapower} model. The power flow characteristics are assumed to remain unchanged unless there is any modification to the power network in the subsequent time-steps in the simulation.

\begin{algorithm}[htbp]
\caption{Pseudo-code for simulating interdependent effects in water and power networks and calculating resilience metrics}
\label{alg:Method1}
\begin{algorithmic}\small
    \State timeList[] = sorted array of unique time stamps in the event table
    \For {$i = 0$ to $timeList.length$ - 1}
        \State Update operating status of the failed components at timeList[i] in both the power and water network models
        \State Run the power network model and calculate the network performance at timeList[i]
        \State Update the indirectly affected components in the water network if any based on the predefined dependencies
        \State Run the water network model for the time interval timeList[i] and timeList[i+1] and calculate the network performance
        \State Update the indirectly affected components in the power network if any based on the predefined dependencies
    \EndFor
    \State Calculate the resilience metrics from the network performance time series
    \end{algorithmic}
\end{algorithm}

\subsubsection{Resilience quantification}
The network-wide effects of infrastructure disruptions and the subsequent repair actions are measured using inbuilt resilience metrics. However, the user can define the resilience metric by taking a wide range of considerations other than resilience, such as economic impact, equity, and sustainability. The repair actions are reflected in the resilience metric, which in turn can be used to identify the best resilience strategy or intervention. 

Currently, the model has two measures of performance (MOP), namely, equitable consumer serviceability (ECS) and prioritized consumer serviceability (PCS), to quantify the system- and network steady-state performances. These MOPs are used as the basis for defining the resilience metrics.

Consider an interdependent infrastructure network $\mathbb{K}$ consisting of a set of infrastructure systems denoted by $K: K\in \mathbb{K}$. There are $N$ consumers who are connected to $\mathbb{K}$ and the resource supply from a system $K$ to consumer $i\in N$  at time $t$ under normal operating conditions is represented by $S_{i}^{K} (t)$. 

The ECS approach assumes equal importance to all the consumers dependent on the network irrespective of the quantity of resources consumed from the network. For an infrastructure system, the ECS at time $t$ is given by Equation~\ref{eq:ecs}.

\begin{equation}
\text{ECS}_{K}(t) = \left (\sum_{\forall i: S_{i}^{K}(t) > 0}\frac{s_{i}^{K}(t)}{S_{i}^{K}(t)}  \right )/n_{K}(t) \quad\text{~s.t.~} s_{i}^{K}(t) \leq S_{i}^{K}(t)
\label{eq:ecs}
\end{equation}
where $s_{i}$ is the resource supply at time $t$ under stressed network conditions and $n_{K}(t)$ is the number of consumers with a non-zero normal demand at time $t$. 

In the case of PCS, the consumers are weighted by the quantity of resources drawn by them. This approach assumes that disruptions to serviceability of large-scale consumers, such as manufacturing sector, have larger effect to the whole region compared to small-scale consumers such as residential buildings. The PCS metric of an infrastructure system at time $t$ is given by the Equation~\ref{eq:pcs}.

\begin{equation}
\text{PCS}_{K}(t) = \left (\frac{\sum_{\forall i: S_{i}^{K}(t) > 0} s_{i}^{K}(t)}{\sum_{\forall i: S_{i}^{K}(t) > 0}S_{i}^{K}(t)}  \right ) \quad\text{~s.t.~} s_{i}^{K}(t) \leq S_{i}^{K}(t)
\label{eq:pcs}
\end{equation}

The normal serviceability component ($S_{i}^{K}(t)$) makes both ECS and PCS metrics unaffected by the intrinsic design inefficiencies as well as the temporal fluctuations in demand.

For water distribution networks, pressure-driven approach is chosen as it is reported to be most ideal for the hydraulic simulation under pressure deficient situations. The component resource supply values for water networks are computed as in Equations~\ref{eq:water_t_pda}--\ref{eq:water_base_pda}.

\begin{eqnarray}
    \label{eq:water_t_pda}
    s_{i}^{water}(t) = Q_{i}(t)\\
    \label{eq:water_base_pda}
    S_{i}^{water}(t) = Q_{i}^{0}(t)
\end{eqnarray}
where $Q_{i}(t)$ and $Q_{i}^{0}(t)$ are the water supplied to consumer $i$ during stressed and normal network conditions, respectively.

For power networks, the power supplied to components under normal and stressed network conditions can be calculated using Equations~\ref{eq:power_t}--\ref{eq:power_base}.

\begin{eqnarray}
    \label{eq:power_t}
    s_{i}^{power}(t) = p_{i}(t)\\
    \label{eq:power_base}
    S_{i}^{power}(t) = p_{i}^{0}(t)
\end{eqnarray}
where $p_{i}(t)$ and $p_{i}^{0}(t)$ are the power supplied to consumer $i$ under stressed and normal power network conditions.

The ECS and PCS time series can be used to profile the effect of the disruption on any of the infrastructure systems. To quantify the system-level cumulative performance loss, a resilience metric called Equivalent Outage Hours (EOH), based on the well-known concept of `resilience triangle' \cite{Bruneau2003}, is introduced. EOH of an infrastructure system due to disaster event is calculated as in Equation~\ref{eq:system_eoh}.

\begin{equation}
    EOH^{K} = \frac{1}{3600}\int_{T_{0}}^{T} \left [ 1 - PCS_{K}(t)\right ]dt \quad \text{or}\quad EOH^{K} = \frac{1}{3600}\int_{T_{0}}^{T} \left [ 1 - ECS_{K}(t)\right ]dt
    \label{eq:system_eoh}
\end{equation}
where $T_{0}$ is the time of the disaster event in the simulation and $T$ is the maximum simulation time (both in seconds). In Equation~\ref{eq:system_eoh}, system performance during normal operating conditions is 1 due to the expression of the MOP used (see Equations ~\ref{eq:ecs}-~\ref{eq:pcs}).

EOH of an infrastructure system can be interpreted as the duration (in hours) of a full infrastructure service outage that would result in an equivalent quantity of reduced consumption of the same service by all consumers during a disaster. The larger the EOH value, the larger the impact on the infrastructure network and thereby on the consumers due to the disruptive event. The EOH metric can effectively capture the response and resilience of the infrastructure system (Equation~\ref{eq:system_eoh}), according to the serviceability criteria chosen by the user. 

Similar to EOH of a system, the consumer-level EOH can also be quantified, which indicates the equivalent duration of infrastructure service outage experienced by each consumer (Equation~\ref{eq:consum_eoh}).

\begin{equation}
    EOH_{i}^{K} = \int_{T_{0}}^{T} \left [ 1 -\frac{s_{i}^{K}(t)}{S_{i}^{K}(t)}\right ]dt
    \label{eq:consum_eoh}
\end{equation}

 Finally, in order to compute the resilience of the interdependent infrastructure network, a weighted EOH metric is derived (Equation~\ref{eq:wEOH}).

\begin{equation}
\overline{\text{EOH}} = \sum_{K\in \mathbb{K}}w_{K}\text{EOH}_{K}
\label{eq:wEOH}
\end{equation}
By default, equal weights are applied to both water and power networks.

\subsection{Experimental Testbeds}
The InfraRisk package is equipped with two self-standing experimental testbeds to carry out generic infrastructure resilience studies, namely, the simple network and the modified Micropolis network. Both networks consist of water-, power-, and traffic networks, and their functional interdependencies.

\subsubsection{Simple network}
As the name indicates, the simple integrated network is developed by combining relatively smaller power-, water-, and traffic networks. The networks are constructed using the most basic infrastructure components. The power network consists of three loads, one external grid connection, and five powerlines. The water network consists of 12 pipelines, nine demand nodes, one water pump, and a tank. The traffic network consists of 22 road links, and nine trip generator- and attractor zones. The details related to the infrastructure components in the integrated network are presented in Table~\ref{tab:simplenet}.
\begin{table}[htbp]\small
\caption{Component details of the simple network}
\begin{tabular}{p{2.5cm}Q{1.25cm}p{2.5cm}Q{1.25cm}p{2.75cm}Q{1.25cm}}
\toprule
\multicolumn{2}{c}{Power network} & \multicolumn{2}{c}{Water network}           & \multicolumn{2}{c}{Traffic network} \\
\midrule
Component            & Count      & Component            & Count      & Component      & Count    \\
\midrule
Buses/poles & 9  & Main pipelines       & 12 & Arterials& 22       \\
Loads& 3  & Demand nodes & 9  & Attractor zones     & 9\\
Motors      & 1  & Pumps& 1  & Generator zones     & 9\\
Switches     & 0  & Tanks& 1  &  &  \\
External grids       & 1  & Reservoirs   & 1  &  &  \\
Power lines  & 5  &      &    &  &  \\
Transformers & 2  &      &    &  &   \\
\bottomrule
\end{tabular}
\label{tab:simplenet}
\end{table}

\subsubsection{Modified Micropolis network}
\begin{figure}[htbp]
  \centering
  \begin{subfigure}[t]{10cm}
    \centering
    \fbox{\includegraphics[width=9cm, trim = {0cm 1cm 1cm 1cm}, clip]{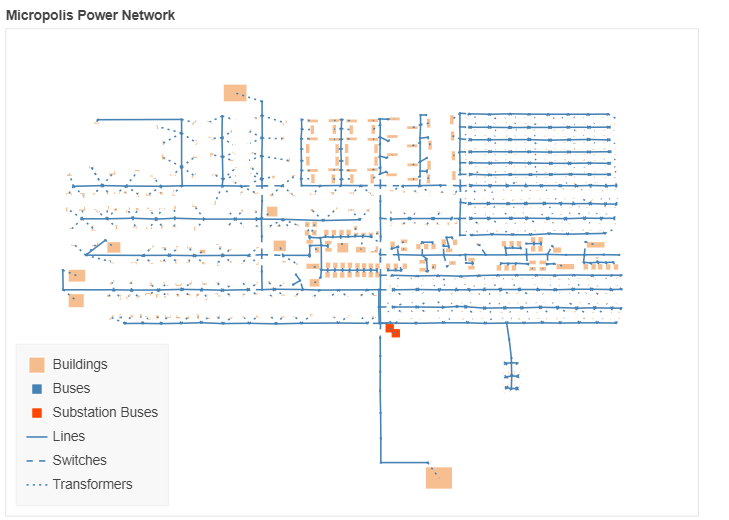}}
    \caption{Power network}
  \end{subfigure}
  \begin{subfigure}[t]{10cm}
    \centering
    \fbox{\includegraphics[width=9cm, trim = {0cm 1cm 1cm 1cm}, clip]{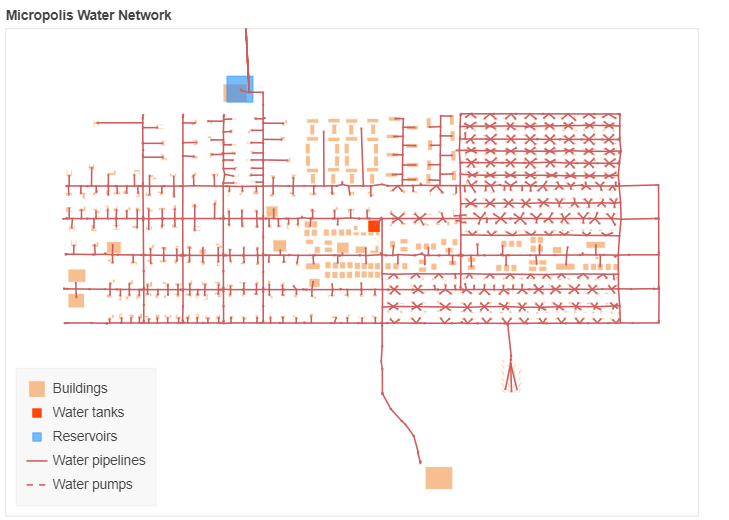}}
    \caption{Water network}
  \end{subfigure}
  \begin{subfigure}[t]{10cm}
    \centering
    \fbox{\includegraphics[width=9cm, trim = {0cm 1cm 1cm 1cm}, clip]{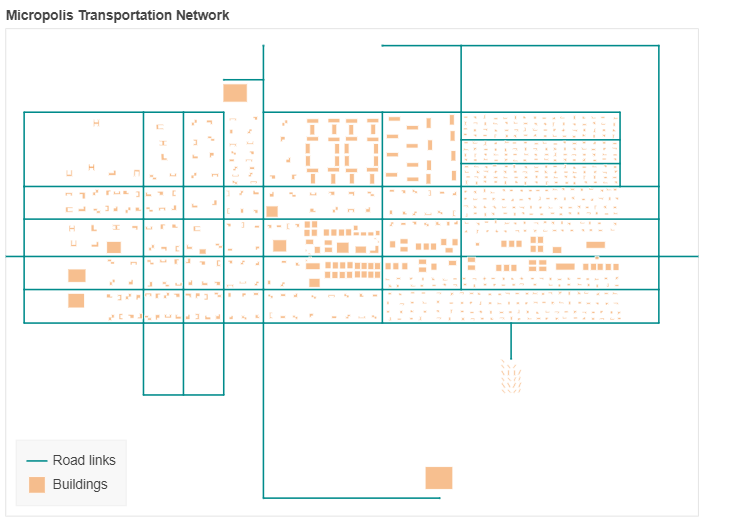}}
    \caption{Road network}
  \end{subfigure}
  \caption{Topology details of the Micropolis integrated power-water-transportation testbed}
  \label{fig:micropolis_testbed}
\end{figure}

The second integrated infrastructure test-bed is based on the well-known Micropolis virtual city developed at the Texas A\&M University. It is a hypothetical city of 5000 inhabitants and is equipped with residential areas, industrial units, and a central business district. Micropolis was originally developed for the design and simulation of water distribution systems \cite{Brumbelow2007} due to the unavailability of real network data. Later its scope was extended to other infrastructure networks, including power networks and communications \cite{Bagchi2009,Ciornei2017} to investigate the role of interdependencies among infrastructure systems. Currently, the testbed consists of water-, power, and transportation networks (Figure~\ref{fig:micropolis_testbed}). The asset-level details of the network are presented in Table~\ref{tab:micropolisnet}.

\begin{table}[htbp]\small
\caption{Component details of modified Micropolis network}
\begin{tabular}{p{2.5cm}Q{1.25cm}p{2.5cm}Q{1.25cm}p{2.75cm}Q{1.25cm}}
\toprule
\multicolumn{2}{c}{Power network} & \multicolumn{2}{c}{Water network}   & \multicolumn{2}{c}{Traffic network} \\
\midrule
Component     & Count    & Component  & Count & Component       & Count    \\
\midrule
Buses/poles  & 1225     & Main pipelines      & 678   & Arterials& 24       \\
Loads  & 788      & Building connections & 685   & Subarterials     & 25       \\
Motors& 3& Hydrants      & 52    & Collector streets& 56       \\
Switches & 61 & Valves (modeled as pipes) & 196   & Attractor zones     & 39       \\
External grids & 1& Pumps   & 8     & Generator zones     & 29       \\
Generators     & 0& Tanks   & 1     &  &  \\
Power lines    & 378      & Reservoirs  & 2     &  &  \\
Transformers   & 789      & &       &  &       \\  
\bottomrule
\end{tabular}
\label{tab:micropolisnet}
\end{table}

\section{Experiment Simulation}
In this section, an experimental simulation to demonstrate the scope of the InfraRisk package is presented. The simulation is performed on the modified Micropolis network (Figure~\ref{fig:micropolis_testbed}). The objective of the experimental simulation is to quantify the direct and indirect component-level operational risks to infrastructure systems resulting from hypothetical floods along a stream in the Micropolis city. The direct risks are evaluated using the hazard initiation module, whereas the interdependent simulations are performed to derive the indirect risks. Statistical analysis is also performed to determine the effectiveness of various heuristics-based resilience strategies.

\subsection{Hypothetical floods and direct impacts on infrastructure components}
\begin{figure}[htbp]
\centering
    \fbox{\includegraphics[width=12cm, trim = {0 0 1.5cm 0}, clip]{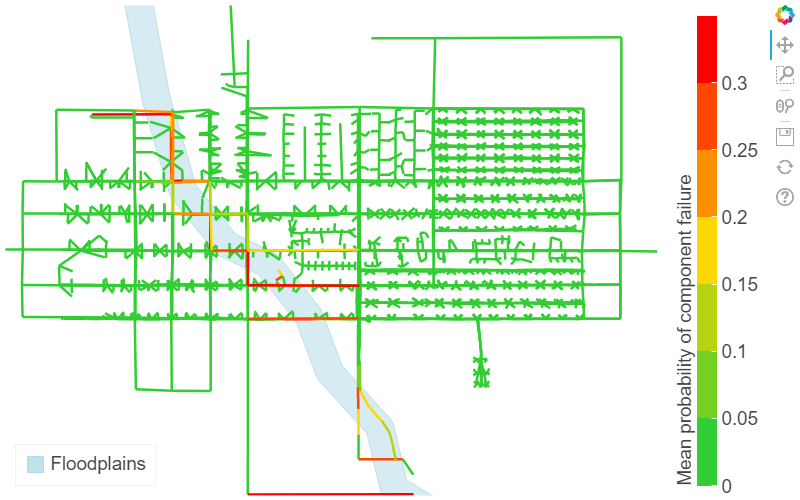}}
    \caption{Probability of failure of infrastructure components (all three infrastructure systems) based on the simulated floods}
    \label{fig:dir_impacts}
\end{figure}

To generate hypothetical floods, a flood hazard profile is required. However, since Micropolis is a virtual city, there is no real historical disaster data to derive the flood risks for the network. Therefore, the authors generated the floods with assumptions on a few of the disaster parameters. It is assumed that the floodplains (regions exposed to floods) are within 100m from the centerline of the main water stream. Three flood intensities are considered, namely, low, moderate, and high with respective probabilities of occurrence of 0.1, 0.3, and 0.5. The conditional probability of exposure ($p(\text{exposure}|\text{hazard})$) is determined based on the distance criterion as explained in subsection~\ref{sec:hazards}. The conditional component failure probabilities ($p(\text{failure}|\text{exposure})$) corresponding to low, moderate, high, and extreme intensity floods generated in the study are 0.1, 0.3, and 0.6, respectively. A total of 165 flood events are simulated, and the corresponding network component failures are determined using Equation~\ref{eq:failureprob}. Only water main pipelines (52 nos), power lines (22 nos), and road links (17 nos) are considered for failure as these are the most critical components of the respective infrastructure systems and are exposed to the simulated floods. For the water pipelines, a failed state represents a leakage with area equivalent to half of the cross-sectional area of respective pipe. In the case of power lines and transportation links, a full failure is assumed. Figure~\ref{fig:dir_impacts} shows the failure probabilities of exposed infrastructure components to the simulated floods.

\subsection{Simulation of interdependent effects}
The next step is to perform the simulation of network-wide impacts of the initial infrastructure disruptions due to each of the flood events. The network-wide impacts are simulated under the three recovery strategies (based on maximum resource-flow handled, betweenness centrality, and land use/zoning). The repair sequences based on each of these strategies are derived by the recovery module and the resultant repair actions are scheduled and implemented in separate simulations. The prioritized component serviceability (PCS) is used as the measure of the performance of systems.

\begin{figure}[htbp]
\centering
  \begin{subfigure}{0.65\textwidth}
    \fbox{\includegraphics[width=1.15\textwidth, trim = {0 0 1.5cm 0}, clip]{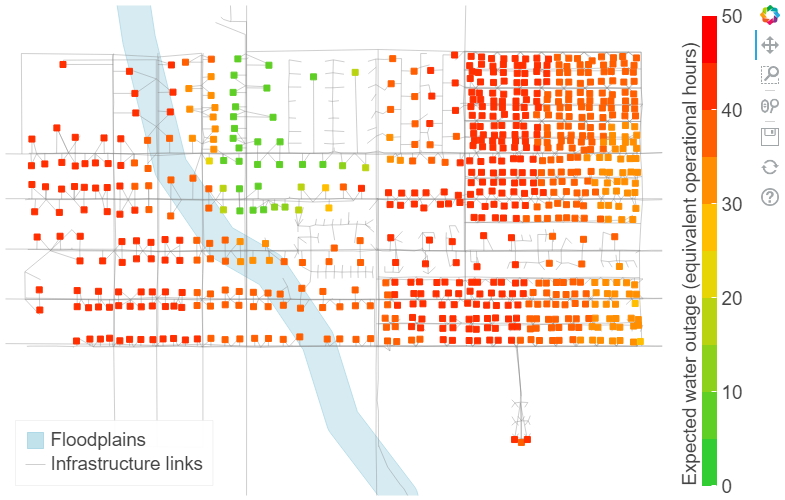}}
    \caption{Expected water disruption in equivalent outage hours {\color{red}(EOH)}}
  \label{fig:water_centrality}
  \end{subfigure}\par
  \begin{subfigure}{0.65\textwidth}
    \fbox{\includegraphics[width=1.15\textwidth, trim = {0 0 1.5cm 0}, clip]{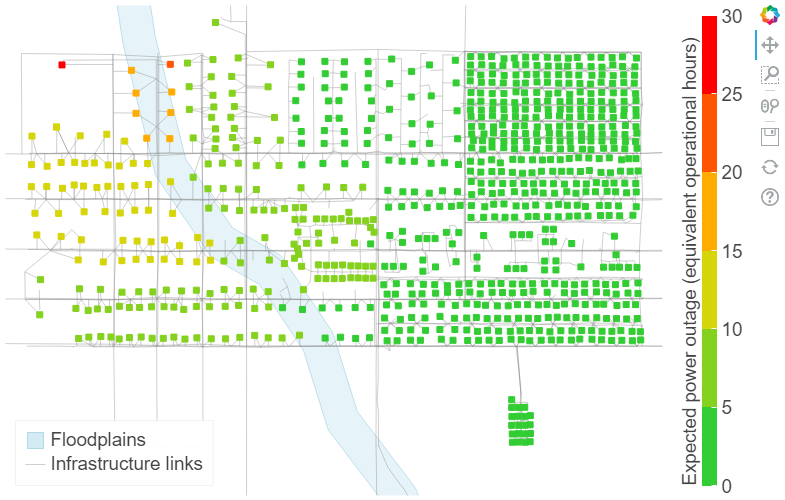}}
    \caption{Expected power disruption in equivalent outage hours {\color{red}(EOH)}}
  \label{fig:power_centrality}
  \end{subfigure}
  \caption{Expected consumer-level impacts for centrality-based repair strategy}
  \label{fig:strategy_compons}
\end{figure}

The consumer-level EOH values are evaluated under various strategies for all the simulated flood scenarios. Figure~\ref{fig:strategy_compons} illustrates the mean EOH of the water and power consumers resulting from the simulated floods when centrality-based restoration strategy was adopted.

\subsection{Analysis of the effectiveness of recovery strategies}
To statistically test if the three recovery strategies vary in terms of their effectiveness in bringing back the integrated network to the pre-disaster state, repeated measures Analysis of Variance (ANOVA) combined with post-hoc tests are employed. Repeated measures ANOVA is a technique used to detect any statistical difference in the means of related (not independent) groups. The results of the repeated measures ANOVA for both water and power networks are presented in Table~\ref{tab:repeated_measures}.
\begin{table}[htbp]\small
\centering
\caption{Repeated measures ANOVA results}
\begin{tabular}{lllllll}
\toprule
Infrastructure & Source & SS & df  & MS  & F  & p-value   \\
\midrule
\multirow{2}{*}{Water} & Strategy & 3096.94 & 2   & 1548.47 & 17.35 & $<$0.001 \\
  & Error & 28910.06 & 324 & 89.22  &  & \\
\midrule
\multirow{2}{*}{Power} & Strategy & 10.86 & 2   & 5.43 & 2.67  & 0.07  \\
  & Error & 656.93 & 324 & 2.02  &  &  \\
\bottomrule
\end{tabular}
\label{tab:repeated_measures}
\end{table}

In the case of the water network, the repeated measures ANOVA rejected the null hypothesis in favor of the alternate hypothesis that the expected EOH significantly differed across the recovery strategies [$F(2, 324)$=17.35, $p < 0.05$]. The results indicate that there is a significant difference in the effectiveness of recovery strategies in minimizing the cumulative disaster impacts on the consumers. However, in the case of power network, the repeated measures ANOVA did not reject the null hypothesis that the expected EOH across strategies are not significantly different. [$F(2, 324)$=2.67, $p > 0.05$].

Motivated by the results of the repeated measures ANOVA tests, post-hoc pairwise t-tests using Benjamini and Hochberg's correction are employed to identify the most-effective recovery strategy for the infrastructure systems (Table~\ref{tab:post-hoc}). In the case of the Micropolis water network, the results show that the expected EOH is the highest for centrality-based strategy (mean = 37.88 hours) and the lowest for capacity-based strategy (mean = 31.80 hours). The expected EOH of zone-based strategy is 35.71 hours. The results suggests the higher effectiveness of capacity-based strategy for Micropolis water network compared to that of centrality- and zone-based strategies. On the other hand, for the power network, the post-hoc tests confirm the findings from the repeated measures ANOVA that there is no statistical difference in the expected EOH among strategies based on capacity (mean = 3.17 hours), centrality (mean = 3.13 hours), and zone (mean = 3.47 hours). The higher mean EOH values of water networks compared to that of power network confirms that the latter is more resilient against the flood impacts.

\begin{table}[htbp]\small
\centering
\caption{Post-hoc tests for pairwise comparison of effectiveness of recovery strategies}
\begin{tabular}{lllrrr}
    \toprule
    Infrastructure & Strategy 1 & Strategy 2 & Mean diff. & T & p-value  \\
    \midrule
    \multirow{2}{*}{Water} & Capacity     & Centrality   & -6.08  & -5.53 & $<$0.0001 \\
    & Capacity     & Zone & -3.91  & -3.69 & $<$0.0001 \\
    & Centrality   & Zone & 2.16   & 2.21 & 0.002 \\
    \midrule
    \multirow{2}{*}{Power} & Capacity     & Centrality   & 0.04 & 0.37       & 0.709\\
    & Capacity     & Zone &  -0.29  & -1.59 & 0.168 \\
    & Centrality   & Zone  & -0.33  & -1.89 & 0.168 \\
    \bottomrule
\end{tabular}
\label{tab:post-hoc}
\end{table}

The results indicate that each infrastructure system may have specific repair strategies that may be the most effective, even when the amounts of resources available for recovery for them are similar. This could be attributed to the unique topological, functional, and recovery characteristics of infrastructure networks.

\section{Conclusions}
This research presented an open-source simulation package for simulating disaster impacts on interconnected power-, water-, and traffic networks. Existing infrastructure-specific simulation models are integrated using a sequential simulation framework for interdependent infrastructure simulations. The details of the constituent modules developed for the integrated infrastructure network modeling, hazard initiation and vulnerability modeling, recovery modeling, simulation of direct and indirect disaster impacts, and resilience quantification, are discussed. Later, the features and capabilities of the simulation package are demonstrated using a case study based on a virtual integrated network. While the infrastructure model can simulate the asset-level performance of the infrastructure systems considering the interdependencies, the resilience indices are designed to track the quality of infrastructure services at the consumer-level. The above features make the simulation platform unique and capable of testing realistic resilience policies, strategies, and interventions in a controlled environment. Some of the potential applications of the package are listed below.
\begin{itemize}
    \item Assessment of pre-disaster asset- and system-level resilience enhancement measures, such as structural enhancements.
    \item Identification of system-level vulnerabilities and critical components.
    \item Post-disaster recovery enhancement by improving existing restoration strategies and protocols, resource allocation, etc.
    \item Resilience-based design of greenfield urban infrastructure networks.
\end{itemize}

While InfraRisk offers a wide-range of applications in infrastructure resilience research, it also has a few shortcomings and limitations, which could be addressed in future research. Some of these limitations are listed below.
\begin{itemize}
    \item Since InfraRisk has integrated existing infrastructure simulation models, the limitations of those models must also be considered in its application. 
    \item The hazard module in InfraRisk currently models exposure, vulnerability, and consequence (failure of components) purely from a probability perspective. These aspects could be better modeled by incorporating disaster-dependent and infrastructure-specific fragility curves, leading to more reliable simulation results.
    \item There could be additional interdependencies, which may be relevant to risk and resilience assessment of infrastructure systems, depending upon the scope of the study.
\end{itemize}

\section*{Acknowledgements}
This research is supported by the National Research Foundation, Prime Minister’s Office, Singapore under its Campus for Research Excellence and Technological Enterprise (CREATE) programme.

The authors extend their gratitude to Prof. Stephen Boyles at the University of Texas at Austin for providing the traffic assignment model.

The authors would like to extend their sincere gratitude to Felix Kottmann, Nazli Yonca Aydin, Ph.D., and Jun Xing Chin, Ph.D., for their suggestions and feedback in developing the InfraRisk package.

\section*{CRediT Author Statement}
\noindent \textbf{Srijith Balakrishnan}: Conceptualization, Methodology, Data curation, Software, Visualization, Formal Analysis, Writing-Original Draft, Project administration. \textbf{Beatrice Cassottana}: Conceptualization, Methodology, Validation, Resources, Writing-Review and Editing, Supervision. 

\section*{Data Availability}
The InfraRisk package can be downloaded from \href{https://github.com/srijithbalakrishnan/dreaminsg-integrated-model}{Github}\footnote{\href{https://github.com/srijithbalakrishnan/dreaminsg-integrated-model}{https://github.com/srijithbalakrishnan/dreaminsg-integrated-model}}. Documentation and codes for sample simulations are available in the InfraRisk package.

\setlength{\bibsep}{0pt plus 0.3ex}
{\small\bibliography{mybibfile}}

\end{document}